\documentclass[A4,10pt]{article}
\usepackage[margin=3cm]{geometry}
\usepackage{palatino}
\usepackage{footnote}
\usepackage{hyperref}
\usepackage{microtype}
\usepackage{xcolor}
\usepackage[utf8]{inputenc}
\usepackage[T1]{fontenc}
\usepackage{natbib}
\definecolor{linkColour}{RGB}{77,71,179}
\hypersetup{
colorlinks=true,
linkcolor=linkColour,
filecolor=linkColour,
urlcolor=linkColour,
citecolor=linkColour,
}

\author{Rendani Mbuvha \textsuperscript{$\psi$}\footnote{Queen Mary University of London {corresponding email: r.mbuvha@qmul.ac.uk}} \and Yassine Yaakoubi\textsuperscript{$\psi$}\footnote{McGill University}\and John Bagiliko \textsuperscript{$\psi$}\footnote{University of Rwanda and African Institute for Mathematical Sciences}\and Santiago Hincapie Potes\textsuperscript{$\psi$} \footnote{Google Research}  \and Amal Nammouchi \textsuperscript{$\psi$}\footnote{Karlstad University} \and Sabrina Amrouche \textsuperscript{$\psi$}\footnote{ZYTLYN \\\textsuperscript{$\psi$} AfriClimate AI}}

\usepackage{parskip}
\usepackage{multirow}
\usepackage{graphicx}
\usepackage{makecell}

\title{Leveraging AI for Climate Resilience in Africa: Challenges, Opportunities, and the Need for Collaboration}

\date{\today}
\begin{document}

\maketitle

\begin{abstract}

As climate change issues become more pressing, their impact in Africa calls for urgent, innovative solutions tailored to the continent's unique challenges. While Artificial Intelligence (AI) emerges as a critical and valuable tool for climate change adaptation and mitigation, its effectiveness and potential are contingent upon overcoming significant challenges such as data scarcity, infrastructure gaps, and limited local AI development. This position paper explores the role of AI in climate change adaptation and mitigation in Africa. It advocates for a collaborative approach to build capacity, develop open-source data repositories, and create context-aware, robust AI-driven climate solutions that are culturally and contextually relevant.

\end{abstract}


\subsection*{AI for Climate Resilience in Africa}

The latest report from the Intergovernmental Panel on Climate Change (IPCC) sixth assessment unequivocally declares that human-induced climate change manifests in many extreme weather and climate phenomena globally~\cite {IPCC_2022_WGIII}. Particularly troubling is the disproportionate impact on vulnerable communities, many of which have contributed minimally to the problem. Regions such as Central, East, and West Africa have emerged as hotspots of vulnerability, where climate-related mortality rates in high vulnerability areas can be as much as 15 times higher than in less vulnerable areas~\citep{IPCC_2022_WGIII}. Moreover, Africa has experienced an accelerated rate of warming, with temperatures rising by +0.3 °C per decade from 1991 to 2022, compared to +0.2 °C per decade from 1961 to 1990 ~\citep{ranasinghe2021climate}. Most African countries are projected to encounter unprecedented high temperatures earlier this century than wealthier, higher-latitude nations. This warming trend has coincided with below-average rainfall across much of the continent, exacerbating drought conditions, particularly in the Horn of Africa, including Ethiopia, Kenya, and Somalia~\citep{wmo2022}.
Meanwhile, other parts of Africa, such as the Sahel and southern regions, have faced the opposite extreme, with intense rainfall events leading to catastrophic flooding. Africa's livelihoods and national economies are closely intertwined with climate conditions, with agriculture alone supporting 55\% 

Artificial Intelligence (AI) is rapidly gaining traction as an essential tool for adaptation and mitigation to climate change. AI applications for climate action are diverse, encompassing climate modelling and forecasting, sustainable energy, transportation, and infrastructure development \citep{rolnick2022tackling}. These applications clearly demonstrate AI's ability to enhance climate resilience by improving predictive capabilities and enabling adaptive responses to climate change. However, the effectiveness of AI is highly contingent upon the availability of robust datasets that capture the complexity of local environmental conditions. Additionally, familiarity with both the downstream application domain and specific regional contexts plays a vital role in the modelling stage. 

For example, understanding and modelling bimodal rainfall patterns in East Africa is critical for agricultural planning and water resource management. These patterns have significant implications for the livelihoods of millions of people who depend on rain-fed agriculture and pastoralism~\citep{james2018evaluating}. Similarly, the African Easterly Jet plays a significant role in shaping the climate of West Africa, particularly affecting the monsoon system and consequently agricultural production and food security in the region~\citep{Nicholson2009}. Despite significant advances in climate modelling in recent decades, none of the current suites of General Circulation Models (GCMs) is built by centres on the African continent; thus, representation of the above critical processes is often not a priority ~\citep{james2018evaluating}.

Consequently, the lack of comprehensive and quality-controlled weather station data in Africa limits the ability to evaluate and improve climate models and tailor them to the unique environmental conditions found across the continent~\citep{james2018evaluating}. While covering one-fifth of the world's land area, Africa has the least developed land-based weather station network of any continent, with a density of weather stations that is only one-eighth of the World Meteorological Organization (WMO) requirement ~\citep{WMO2019}. This network has also been declining in terms of quality and number of stations over time ~\citep{WMO2019,dinku2019challenges}.

This lack of data is compounded by the fact that most climate research funding and AI development have historically been concentrated outside of Africa, using non-African datasets that may not be fully applicable or transferable to Africa's environmental and socio-economic conditions~\citep{niang2014africa, dinku2018enhancing}. More specifically, IPCC~\citep{IPCC2014Africa} shows that only 3.8\% of global funding for climate research was directed towards Africa, despite the continent's high vulnerability to the impacts of climate change. Moreover, most of this funding originated outside Africa and was allocated to non-African research institutions, which underscores data availability and research funding as critical limitations.

As highlighted by \cite{Kull2021}, the performance of Numerical Weather Prediction models, which form the basis of modern forecasting, heavily relies on the availability of comprehensive and quality-controlled meteorological observations. The sparsity of weather stations across the continent leads to significant gaps in understanding atmospheric conditions. This, in turn, translates to lower forecast accuracy, particularly for short-range predictions crucial for disaster preparedness and daily decision-making in weather-sensitive sectors like agriculture, energy, and transportation. Consequently, African nations face increased vulnerability to extreme weather events and struggle to optimise resource management and economic productivity, further exacerbating challenges related to sustainable development and climate change adaptation.

\subsection*{Opportunities to Drive AI for a Climate Resilient  Africa }

While there are challenges to using AI for climate action in Africa, there are also opportunities to overcome these barriers. One such opportunity is to leverage grassroots movements for decentralised research and collaboration. These movements facilitate community involvement, garner local knowledge, and promote transdisciplinary collaboration. By integrating grassroots initiatives with traditional agricultural practices and local ecological knowledge, AI for climate action can be effectively implemented.

Grassroots movements have demonstrated success in other areas where Africa is underrepresented in technology. Examples include Masakhane\footnote{https://www.masakhane.io/}, a movement with over 2000 members actively contributing to the advancement of Natural Language Processing by creating and curating novel datasets and benchmarks that incorporate African languages ~\citep{orife2020masakhane}. Another initiative, AfriClimate AI\footnote{https://www.africlimate.ai/}, is a newly formed grassroots movement focused on addressing Climate Action through open-source data curation, capacity building, and collaboration in pan-African AI-based climate research. The support and success of such movements are core to developing context-aware AI-driven solutions across disciplines.

Open-source climate information services (CIS), such as those provided by the European Centre for Medium-Range Weather Forecasts (ECMWF), have been fundamental in driving exciting advances in climate informatics, as recent studies have shown. These advances include the development of state-of-the-art models such as IceNet ~\citep{andersson2021seasonal}, GraphCast ~\citep{lam2022graphcast}, and Neural GCM ~\citep{kochkov2023neural}.

Numerous opportunities arise in the  African CIS landscape for collaborative continental efforts. These opportunities include building and strengthening weather and climate observation infrastructure and networks for medium and long-term planning, improving the uptake and effectiveness of CIS through coordinated implementation and sustainability measures, and building capacity in the generation, uptake, and effective use of CIS among various stakeholders ~\citep{africanUnion2023}.

In conclusion, we advocate for the creation of a Pan-African roadmap for AI and Climate Science. This roadmap should streamline collaboration among researchers, practitioners, policymakers, and grassroots communities to tackle the aforementioned challenges. This furthers the proposition by \cite{james2018evaluating}, which stresses the need to include African experts in climate model development. It should recognise the opportunity that AI offers to craft context-aware solutions tailored to the continent's specific needs and challenges. In addition to placing a significant emphasis on developing and implementing context-aware climate information services that are fit for purpose, this roadmap is envisaged to cultivate an environment conducive to implementing mitigation and adaptation policies that promote a sustainable, prosperous, and climate-resilient Africa.

\clearpage
\bibliographystyle{apalike}
\bibliography{references}
\end{document}